\documentstyle[12pt,psfig,fig_tab]{article}

\title{
{\normalsize\rm \vspace*{-1.5cm} \hfill Preprint LBNL-39895 \vspace*{+1.0cm}}\\
$^3$He- and $^4$He-induced nuclear fission --\\
a test of the transition state method\footnote{Talk presented at the XXXV. 
International Winter Meeting on Nuclear Physics, Bormio, Italy, 1997
and to be published in the proceedings.}
} 
\author{
\underline{Th. Rubehn},\thanks{Electronic address: TRubehn@lbl.gov}
K. X. Jing, L. G. Moretto,\\
L. Phair, K. Tso, and G. J. Wozniak \\
\vspace*{-2mm}\normalsize Nuclear Science Division,\\
\vspace*{-2mm}\normalsize Lawrence Berkeley National Laboratory,\\
\vspace*{-2mm}\normalsize University of California, Berkeley, 
California 94720, USA
}
\date{ }
\begin{document}
\maketitle
\begin{abstract}
Fission in $^3$He and $^4$He induced reactions
at excitation energies between the fission barrier and 140 MeV
has been investigated. Twentythree
fission excitation functions of various compound nuclei in
different mass regions are shown to scale exactly according to
the transition state prediction once the shell effects are accounted for.
New precise measurements of excitation functions in a mass region
where shell effects are very strong, allow one to
test the predictions with an even higher accuracy.
The fact that no deviations from the transition state method have
been observed within the experimentally investigated excitation
energy regime allows one to assign limits for the fission transient
time.
The precise measurement of fission excitation functions of 
neighboring isotopes enables us to experimentally estimate the first chance 
fission probability. Even if only first chance fission is 
investigated, no evidence for fission transient times larger than 30 zs
can be found.
\end{abstract}

\section{Introduction}
The study of the fission process is -- even more than half a century
after its discovery -- still of general interest. 
This is caused by both the complexity of the process itself and
the availability of new accelerators and techniques that enable
the study of new aspects of fission.
While several interesting experiments have been performed using
relativistic heavy ion beams, we will in this paper concentrate
on the investigation of light particle induced nuclear fission
at excitation energies between the fission barrier and 
$\sim$140 MeV.  
It has been shown recently that a novel analysis 
\cite{Mor95_prl,Rub96_prc,Rub96_snowbird}
allows for the model-independent extraction of fundamental quantities
of the fission process, like effective fission barriers,
shell effects, and the much discussed fission delay time
\cite{Mor95_prl,Hil92,Pau94}.

Fission excitation functions vary dramatically
from nucleus to nucleus over the periodic table
\cite{Rai67,Mor72,Kho66}:
Some of the differences can be understood in terms of a changing
liquid-drop fission barrier with the fissility parameter,
others are due to to strong shell
effects which occur e.g. in the neighborhood of the double magic
numbers $Z$=82 and $N$=126.
Further effects may be associated with pairing and the angular
momentum dependence of the fission barrier \cite{Van73,Wag91}.
                                     
Fission rates have been calculated most often on the basis of
the transition state method introduced by
Wigner \cite{Wig38}, and later applied to fission by
Bohr and Wheeler \cite{Boh39}.
Recent publications claim the failure of the transition state 
rates to account for the measured amounts of prescission neutrons or 
$\gamma$ rays in relatively heavy fissioning systems
\cite{Hil92,Pau94,Tho93}.
This alleged failure has been attributed to the transient time
necessary for the so-called slow fission mode to attain its
stationary decay rate
\cite{Gra83a,Gra83b,Wei84,Gra86,Lu86,Lu90,Cha92,Fro93,Siw95}.
The larger this fission delay time, the more favorably
neutron decay competes with the fission process.
This leads to an effective fission probability smaller
than predicted by the Bohr - Wheeler formula.
The experimental methods of these studies, however, suffer from two
difficulties: First they require a possibly large correction
for post-saddle, but pre-scission emission; second, they
are indirect methods since they do not directly determine
the fission probability.
The measured prescission particles can be emitted either
before the system reaches the saddle point, or during
the descent from saddle to scission. Only from the
anomalies in the first component, would deviations of
the fission rate from its transition state value be expected.
The experimental separation of the two contributions, however, 
is fraught with difficulties which make the evidence ambiguous.
As a different ansatz, we will, in this paper, therefore study 
the validity of the transition state method by directly measuring the
fission probability and its energy dependence over a broad
energy range. By investigating several old and new data sets,
we are able to test the transition state rates for
a large number of systems.

\section{Experiment}
A set of experiments investigating fission excitation
functions of various compound nuclei has been performed 
in the 1960s at Berkeley \cite{Kho66}. All experiments
have used mica detectors and have thus required very high
beam currents and rather long irradiation times to compensate
for the very small angular coverage of the detectors. 
The data have initially been used for a first test of 
the method proposed by Moretto {\it et al.} \cite{Mor95_prl}.

In two recent experiments, we have measured fission excitation 
functions of the compound 
nuclei $^{200}$Tl, $^{211}$Po, $^{212}$At,
and $^{209, 210, 211, 212}$Po, formed in the reactions
$^3$He + $^{197}$Au, $^{206,207,208}$Pb, $^{209}$Bi and
$^4$He + $^{206,207,208}$Pb, respectively.
Both runs have been performed at the Lawrence Berkeley 
National Laboratory's 88 inch cyclotron which delivered
between 19 and 26 different energies per ion species.

\begin{figure}[p]
 \centerline{\psfig{file=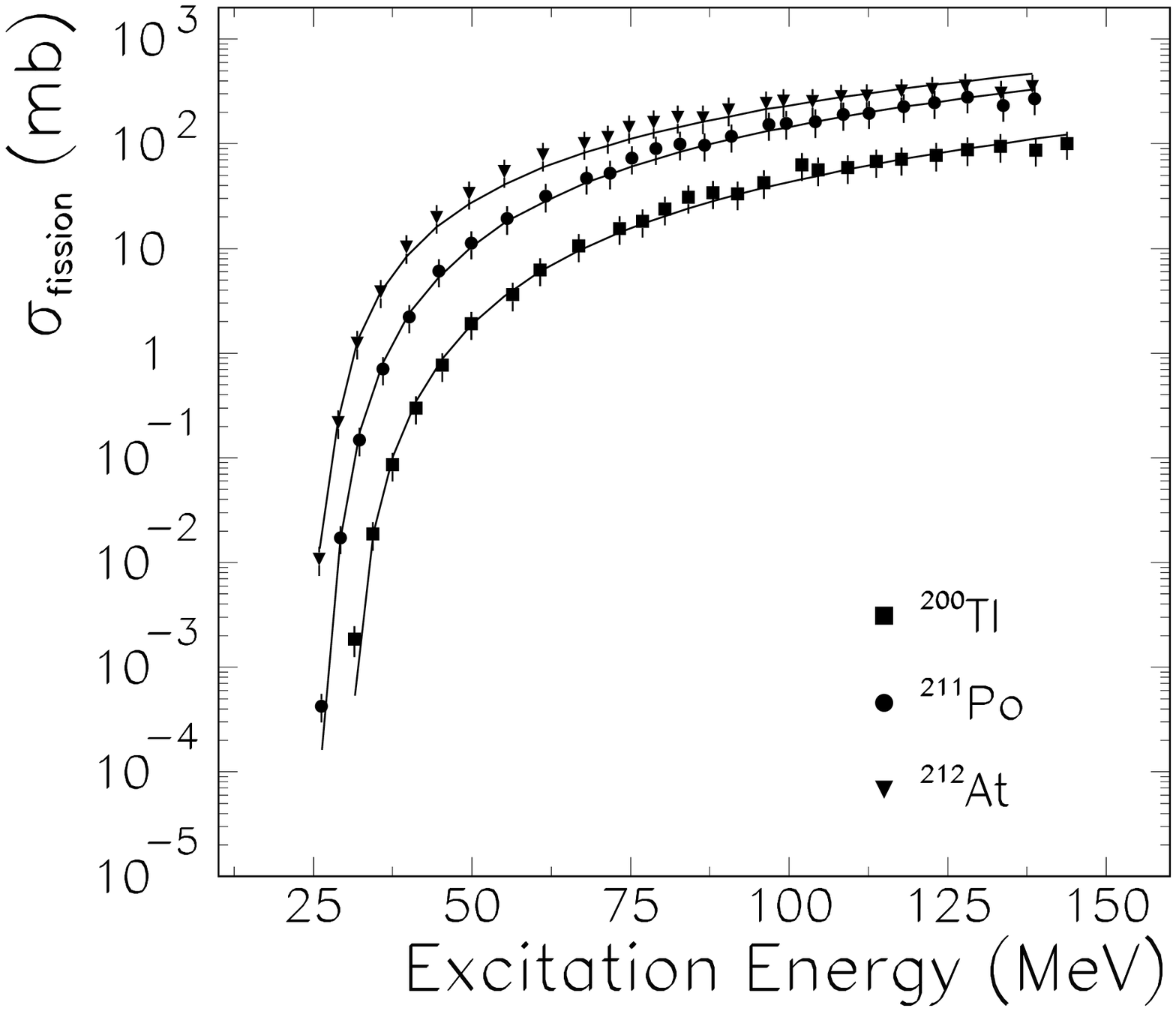,height=7cm}}
 \caption[]{Excitation function for fission of several compound
 nuclei formed in $^{3}$He induced reactions. The
 different symbols correspond to the experimental data points.
 The solid line shows the results of a fit to the data using
 a level density parameter $a_{n} = A/8$.
 The error bars denote the statistical and systematic errors
 combined in quadrature.
 }
 \label{exc_ftn}
 \vspace*{5mm}
 \centerline{\psfig{file=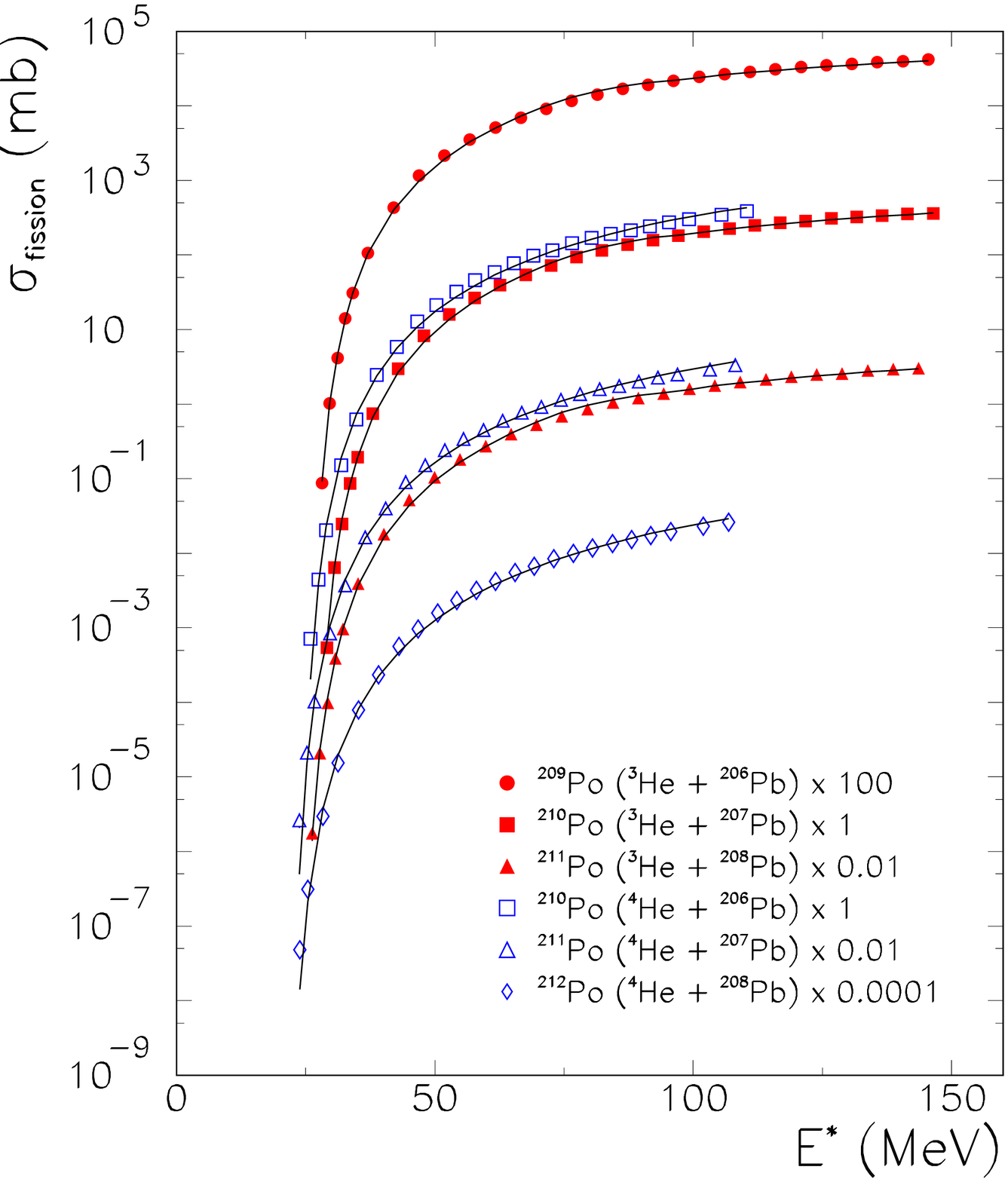,height=7cm}}
 \caption[]{Same as Fig.~\protect\ref{exc_ftn} for fission of the compound
 nuclei $^{209,210,211,212}$Po formed in $^{3}$He and $^{4}$He induced 
 reactions.
 }
 \label{exc_ftn_new}
\end{figure}
 
To cover a large solid angle and, therefore, to minimize
beam time, we performed an experiment using two
large area parallel-plate-avalanche counters (PPACs)
with an active area of 200 x 240 mm$^2$ each. The detectors
were mounted at 80$^{\circ}$ and 260$^{\circ}$ with respect 
to the beam axis, allowing for the detection of both fission 
fragments in coincidence.
The PPACs were placed at a distances of 150~mm from the target
to the center of each detector resulting in a coverage of
approximately 20\% of 4$\pi$.
The accuracy achieved in these experiments is significantly better
than the one of the old runs using mica detectors.

In Fig.~\ref{exc_ftn}, we show the experimental
fission cross sections for the three compound nuclei
$^{200}$Tl, $^{211}$Po, and $^{212}$At as a function of
excitation energy.
The error bars denote both the statistical and the systematic
errors. While the statistical errors dominates at the lowest
energy points, the systematic uncertainties are the main
contribution at higher excitation energies.
Fig.~\ref{exc_ftn_new} shows the results obtained in a
most recent experiment. Here, fission cross
sections of the compound nuclei $^{209,210,211,212}$Po
have been measured in $^3$He and $^4$He induced reactions.
The maximum excitation energies are dominated by the
$K$ of the cyclotron and the significantly different
$Q$ value of the compound nuclei for both projectiles.
The observed systematically higher fission probability 
for the $\alpha$ induced reactions is due to the higher
angular momentum of the compound nuclei. 
The comparison between the data shown in Fig.~\ref{exc_ftn}
and those presented in Fig.~\ref{exc_ftn_new} demonstrates
the improvement made in the quality of the measurements
\cite{Rub96_rbs}.

Cross sections were determined for these fission events using
\begin{equation}
 \sigma_{f} = \frac{n_{f} A}{n_{beam} N_A m} \eta(\theta, \phi),
\end{equation}
where $n_f$ and $n_{beam}$ are the number of fission events and
the number of beam particles, respectively.
$A$ represents the mass number of the target, $N_A$ Avogadro's
constant, and $m$ the thickness of the target. Due to the
incomplete angular coverage, the quantity
$\eta(\theta,\phi)$ which accounts for the geometrical
acceptance and for the non-isotropic emission of the fission fragments
has be be taken into account.
The anisotropic angular distribution
$\frac{(d\sigma / d\Omega)_{\theta}}{(d\sigma / d\Omega)_{90^{\circ}}}$
of the fission fragments has been shown to be reasonably described
by the function $\sin^{-1} \theta$ \cite{Van73}.
We have used this dependence for the determination of our
acceptance. Comparison between our $\alpha$ induced fission excitation
functions and those measuring the angular dependence \cite{Kho66}
agree very well.

The excitation energy was calculated assuming a full momentum and mass 
transfer of the helium ions to the compound nucleus (CN). The binding 
energies of $^3$He, the target isotopes, and the compound nuclei
were taken from nuclear mass tables \cite{Aud93}.

\section{Analysis}
\label{analysis}
The experimental data shown above are analyzed according to 
a rather simple method proposed by Moretto {\it et al.} \cite{Mor95_prl}.
In this section, we briefly reflect the procedure.

Using the transition state expression for the fission decay width 
\begin{equation}
 \Gamma_f \approx \frac{T_s}{2\pi}
 \frac{\rho_s(E - B_f - E^s_r)}{\rho_n(E - E_r^{gs})},
\end{equation}
the fission cross section can be written as follows:
\begin{equation}
 \sigma_f = \sigma_0 \frac{\Gamma_f}{\Gamma_{\rm{total}}}
 \approx \sigma_0 \frac{1}{\Gamma_{\rm{total}}}
 \frac{T_s \rho_s (E - B_f - E^s_r)}{2\pi \rho_n (E - E_r^{gs})},
 \label{e1}
\end{equation}
where $\sigma_0$ is the compound nucleus formation cross section,
$\Gamma_f$ is the decay width for fission
and $T_s$ is the energy dependent temperature
at the saddle; $\rho_s$ and $\rho_n$ are the saddle and ground
state level densities, $B_f$ is the fission barrier,
and $E$ the excitation energy. Finally, $E^s_r$
and $E^{gs}_r$ represent the saddle and ground state rotational
energies.

To further evaluate the expression, we make use of 
$\rho(E) \propto \exp(2\sqrt{aE})$ for the level density and rewrite
Eq.~\ref{e1} as:
\begin{equation}
 \ln\Big( \frac{\sigma_f}{\sigma_0} \Gamma_{\rm{total}}
 \frac{2\pi \rho_n (E - E_r^{gs})}{T_s} \Big) =
 2 \sqrt{a_f (E - B_f - E_r^s)}.
 \label{scal}
\end{equation}

Since the neutron width $\Gamma_n$
dominates the total decay width in our
mass and excitation energy regime, we can write:
\begin{equation}
 \Gamma_{\rm{total}}  \approx \Gamma_n \approx K T_n^2
 \frac{\rho_n(E - B_n - E_r^{gs})}{2\pi \rho_n(E - E_r^s)}
\end{equation}
where $B_n$ represents the binding energy of the last neutron,
$T_n$ is the temperature after neutron emission, and
$K = \frac{2 m_n R^2 g'}{\hbar^2}$ with the spin degeneracy
$g'=2$.

The study of the fission process in the lead region
forces us to take strong shell
effects into account. For the fission excitation functions
discussed in this paper, the lowest excitation energies for
the residual nucleus after neutron emission are of the order of
15-20 MeV and therefore high enough to assume the asymptotic
form for the level density which is given below:
%
\begin{equation}
 \rho_n(E-B_n-E_r^{gs}) \propto
 \exp \big(2 \sqrt{a_n(E-B_n-E_r^{gs}-\Delta_{shell})} \big)
\label{rho_n}
\end{equation}
%
where $\Delta_{shell}$ is the ground state shell effect of the
daughter nucleus ($Z,N-1$).
For the level density at a few MeV above the saddle point, we can use
\begin{equation}
 \rho_s(E-B_f-E_r^{s}) \propto
 \exp \big(2 \sqrt{a_f(E-B_f^*-E_r^{s})} \big)
 \label{rho_s}
\end{equation}
since the large saddle deformation implies small shell effects.
Deviations due to pairing, however, may be expected at very
low excitation energies. In Eq.~\ref{rho_s}, we introduced the
quantity $B_f^*$ which represents an effective fission barrier,
or, in other words, the unpaired saddle energy, i.e.
$B_f^* = B_f + 1/2 g \Delta_0^2$ in the case of an even-even
nucleus and $B_f^* = B_f + 1/2 g \Delta_0^2 - \Delta_0$ for
nuclei with odd mass numbers. Here, $\Delta_0$ is the saddle
gap parameter and $g$ the density of doubly degenerate single
particle levels at the saddle.

Finally, the use of Eq.~\ref{rho_n} and Eq.~\ref{rho_s}
for the level densities allows us to study the scaling of the
fission probability as introduced in Eq.~\ref{scal}:
%
\begin{equation}
 \frac{1}{2 \sqrt{a_n}} \ln \Big(\frac{\sigma_f}{\sigma_0}
 \Gamma_{\rm{total}} \frac{2\pi\rho_n(E-E_r^{gs})}{T_s}\Big) =
 \frac{\ln R_f}{2\sqrt{a_n}}
 = \sqrt{\frac{a_f}{a_n}(E - B_f^* - E_r^s)}.
\label{rf_eq}
\end{equation}
%
The values for $B_f^*$, $\Delta_{shell}$, and $a_f/a_n$ using
$a_n = A/8$ can be obtained by a three parameter fit of
the experimental fission excitation functions;
the best results of the fits are shown in Figs.~\ref{exc_ftn}
and \ref{exc_ftn_new}. 
For this procedure, the formation cross sections $\sigma_0$,
which is approximated by the reaction cross section,
and the corresponding values for the maximum angular momentum
$l_{max}$ were taken from an optical model calculation.

At high beam energies per nucleon (in particular for the $^3$He-induced
reactions) there might be a significantly contribution from 
incomplete fusion. We have calculated fusion cross sections
by using the Bass model and have used them to estimate the 
formation cross section $\sigma_0$. The fit parameters, however,
do only change insignificantly.

\section{Results and interpretation}
\begin{figure}[htbp]
 \centerline{\psfig{file=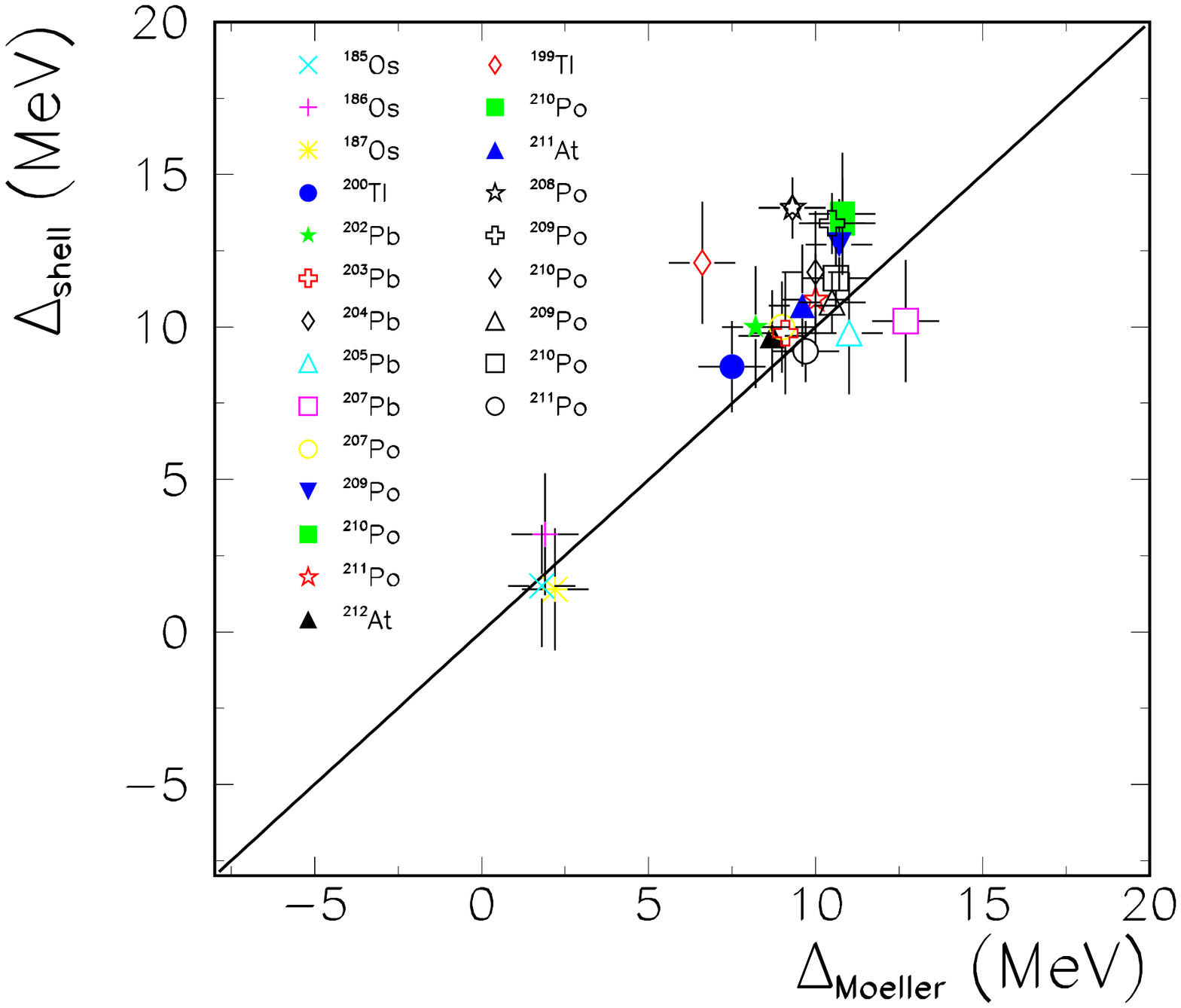,height=7cm}}
 \caption[]{
 Shell corrections for the daughter nuclei, extracted from fits to
 the fission excitation functions plotted against the values 
 determined from the ground state masses.
 }
 \label{delta}
 \vspace*{5mm}
 \centerline{\psfig{file=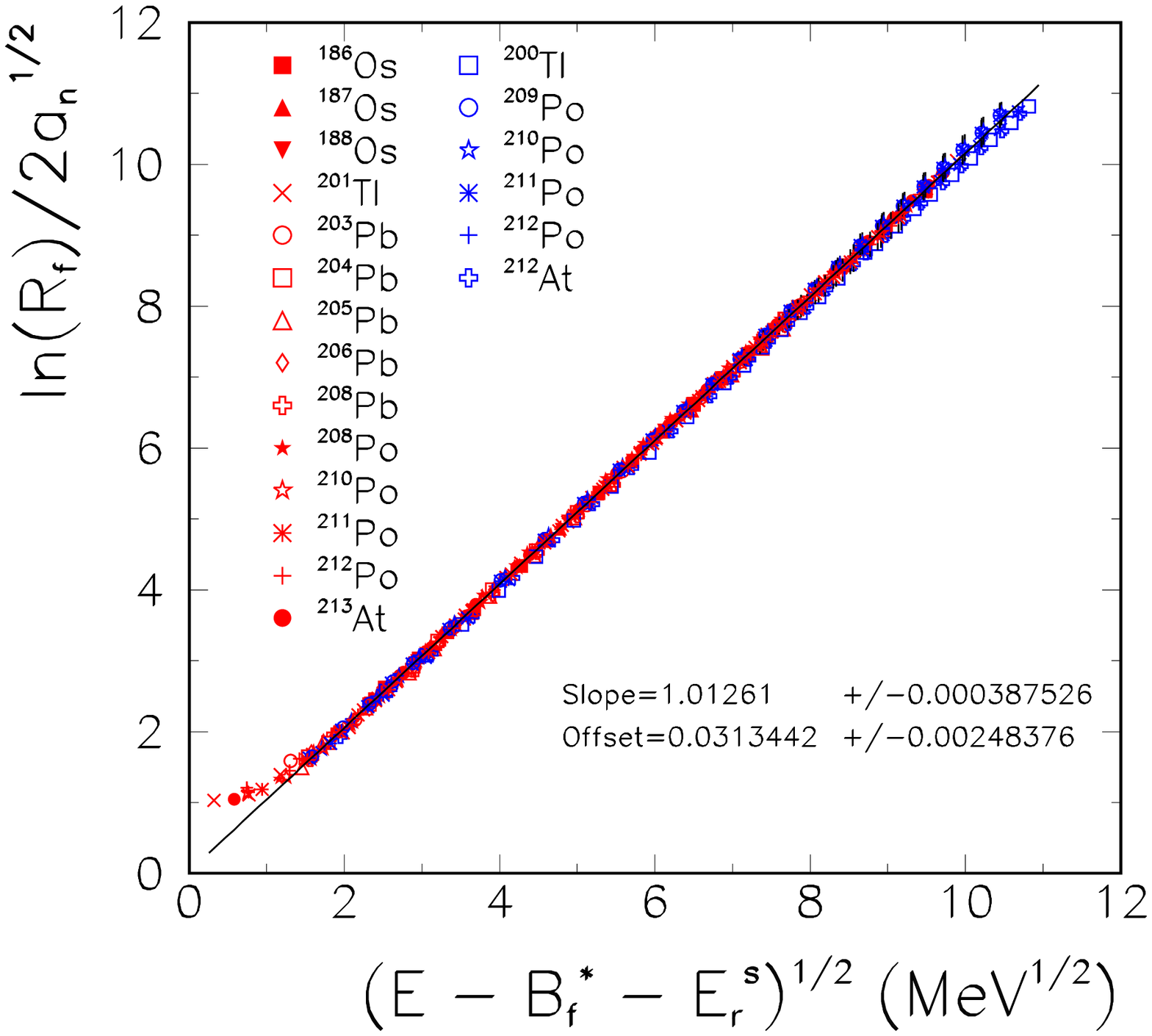,height=7cm}}
 \caption[]{
 The quantity $\frac{\ln R_f}{2 \sqrt{a_n}}$ vs
 the square root of the intrinsic excitation energy over the
 saddle for fission of several compound nuclei as described
 in the text. The straight line represents a fit to the entire
 data set.
 }
 \label{rf}
\end{figure}

In Fig.~\ref{delta}, we show the shell corrections obtained from 
the fit to our experimental data. In addition, we also show the 
results of a similar analysis \cite{Mor95_prl} of fission excitation
functions measured in the 1960's.  The observed correlation is very
good, especially if one reflects how difficult it is to establish
a good liquid drop baseline.
We point out that the method applied here represents a totally
independent way to determine the ground state shell effects.

As pointed out before, plotting the left hand side of Eq.~\ref{scal}
versus the expression $\sqrt{E-B_{\rm f}-E_{\rm r}^{\rm s}}$ should result 
in a straight line if the transition state predictions hold.
In Fig.~\ref{rf}, we show the results for a large number of 
fission excitation functions. 
A remarkable straight line can be observed for all compound nuclei
investigated. It should be noted that the scaling extends over six 
orders of magnitude in the fission probability although shell effects
are very strong for several nuclei. 
A fit to the data results in a straight line that goes through the
origin and has a slope that represents $a_{\rm f}/a_{\rm n}$,
consistent with unity. 
The observed scaling and the lack of deviations indicates that
the transition state rates hold well. 

\begin{figure}[tb]
 \centerline{\psfig{file=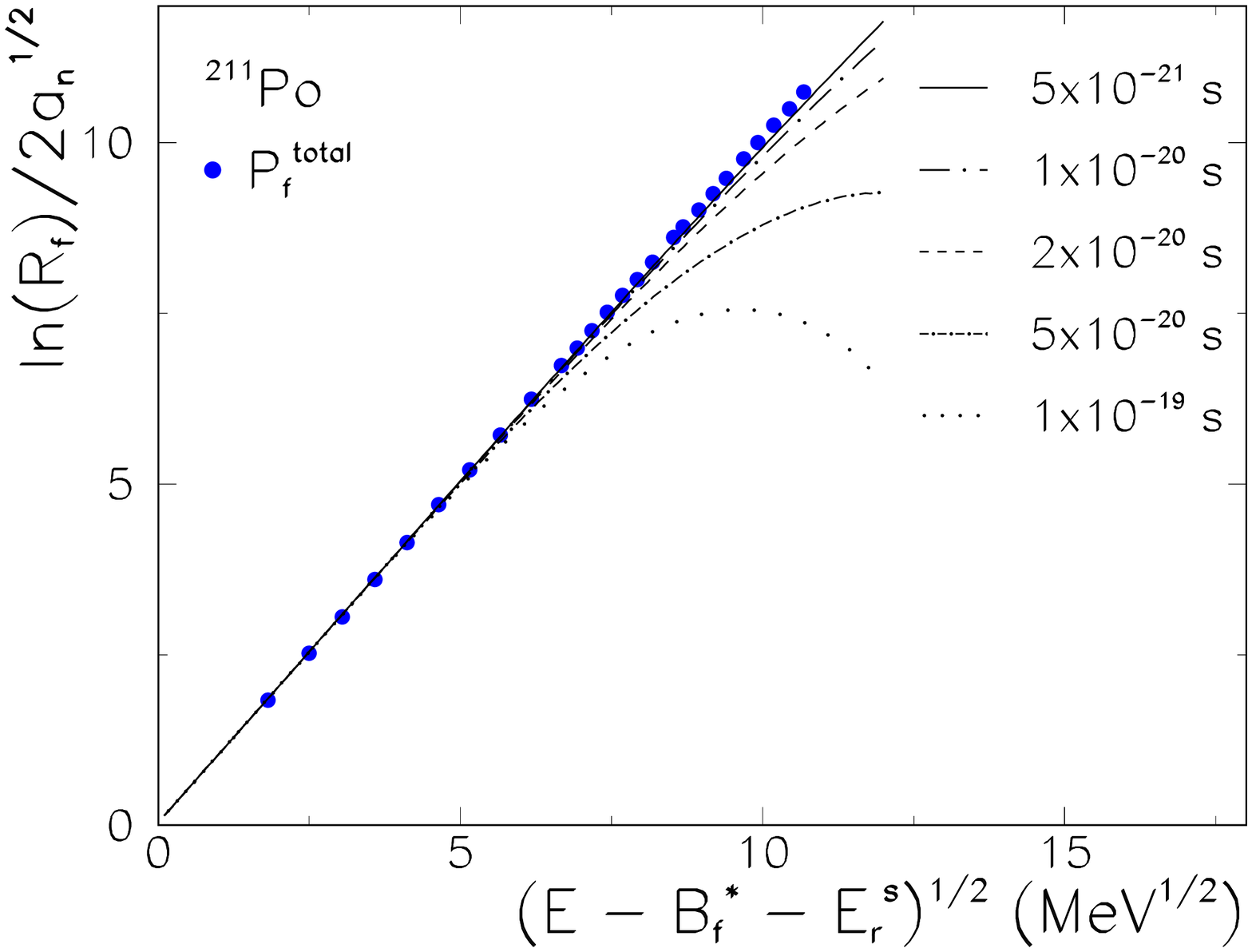,height=7cm}}
 \caption[]{Same as Fig.~\protect\ref{rf} for $^{211}$Po.
 The lines represent calculations assuming that no fission occurs
 during a given transient time which is indicated in the
 figure. For further details see text.
 }
 \label{time}
\end{figure}

The excitation energy range covered by the experiments presented
here correspond to life times of the compound nuclei between
10$^{-18}$ and 10$^{-22}$ s and should therefore be sensitive to
delay times in the fission process. 
To investigate this effect, we assume a step
function for the transient time effects. In this assumption,
the fission width can be written as follows:
\begin{equation}
 \Gamma_f = \Gamma_f^{\infty} \int^{\infty}_{0} \lambda(t)
 \exp(\frac{-t}{\tau_{CN}}) d(\frac{t}{\tau_{CN}}) =
 \Gamma_f^{\infty} \exp(\frac{-\tau_D}{\tau_{CN}})
\end{equation}
where the quantity $\lambda(t)$ jumps from 0 at times smaller
than the transient time $\tau_D$ to 1 for times larger than
$\tau_D$. Furthermore, $\Gamma_f^{\infty}$ denotes the
transition state fission decay width and $\tau_{CN}$
represents the life time of the compound nucleus.
This expression for the fission decay width has been
used in the formalism described above; the parameters
$B_f^*$, $\Delta_{shell}$, and $a_f/a_n$ have been
taken from the fit.
In Fig.~\ref{time}, we show the results of these
calculations for the compound nucleus $^{211}$Po; the
different lines indicate different assumed values of the
transient time between 1$\times$10$^{-19}$ and 5$\times$10$^{-21}$
seconds.
The calculated values show an obvious deviation
from the experimental data for assumed
transient times larger than 10$^{-20}$ seconds.

In the formalism above, we have only accounted for 
first chance fission while for the experimental data,
we have used the total fission probability. 
At low excitation energies,
first chance fission will certainly be the most 
dominant contribution. However, at the highest energies
higher chance fission is expected to become more and
more important.
In the following, we shall thus estimate the percentage of
first chance fission from our experimental data.

To extract first chance fission from experimental data,
we have measured fission excitation functions of 
neighboring Po isotopes. 
The difference in the fission probability determines
the 1st chance fission probability.
Since the energy dependence
of the first chance fission probability is determined
by subtracting similar cross sections of the two 
neighboring isotopes, it is essential to measure the
cumulative cross sections with high precision; we have 
discussed this in detail in Ref.~\cite{Rub96_rbs}. 
The data measured in a recent experiment (see Fig.~\ref{exc_ftn_new})
should be precise enough to determine the 1st chance 
fission probability.

\begin{figure}[tbp]
 \centerline{\psfig{file=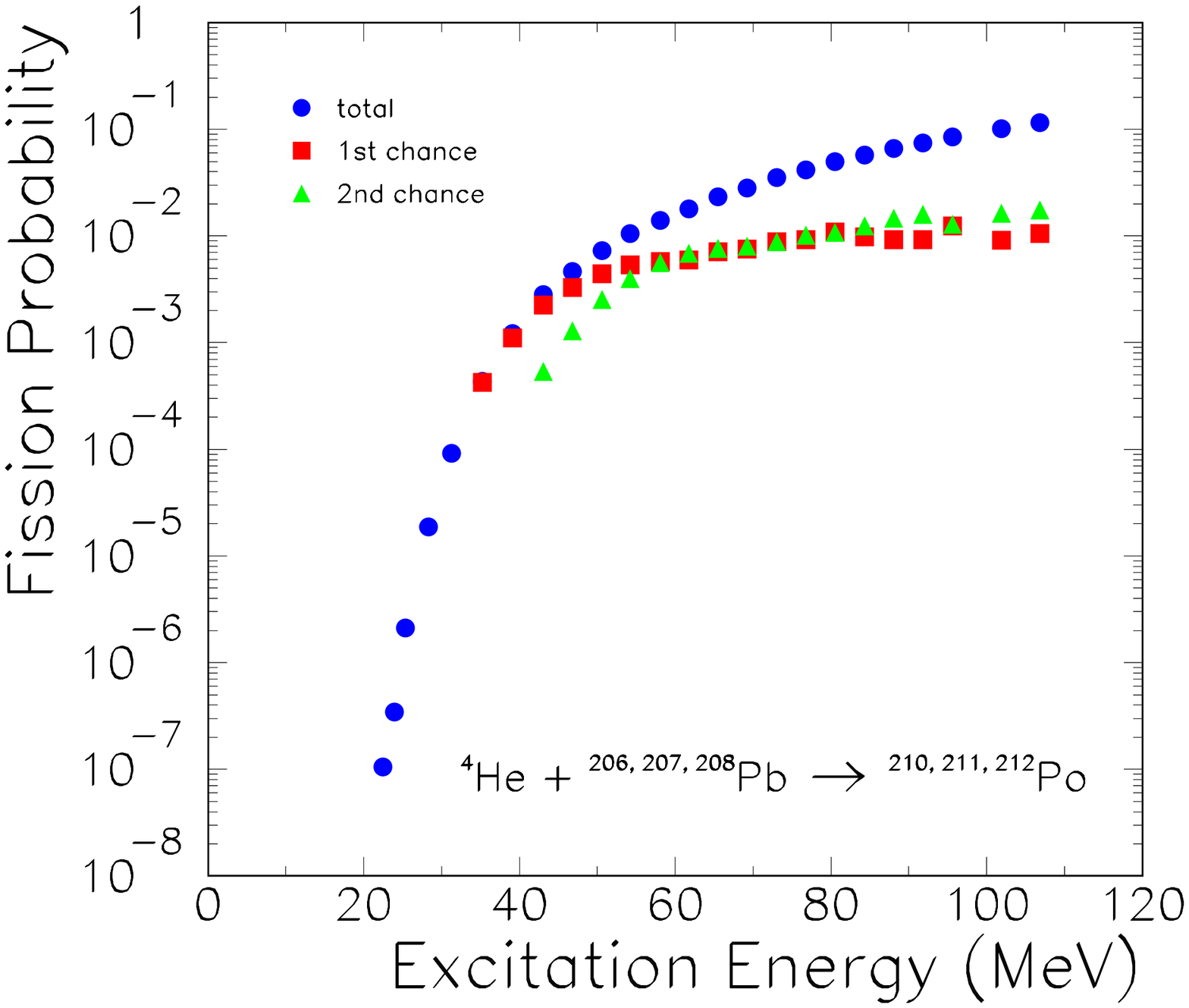,height=7cm}}
 \caption[]{
 {\bf PRELIMINARY} first and second chance fission probability for the 
 reaction $^4$He + $^{206,207,208}$Pb. 
 }
 \label{1stchance}
 \vspace*{5mm}
 \centerline{\psfig{file=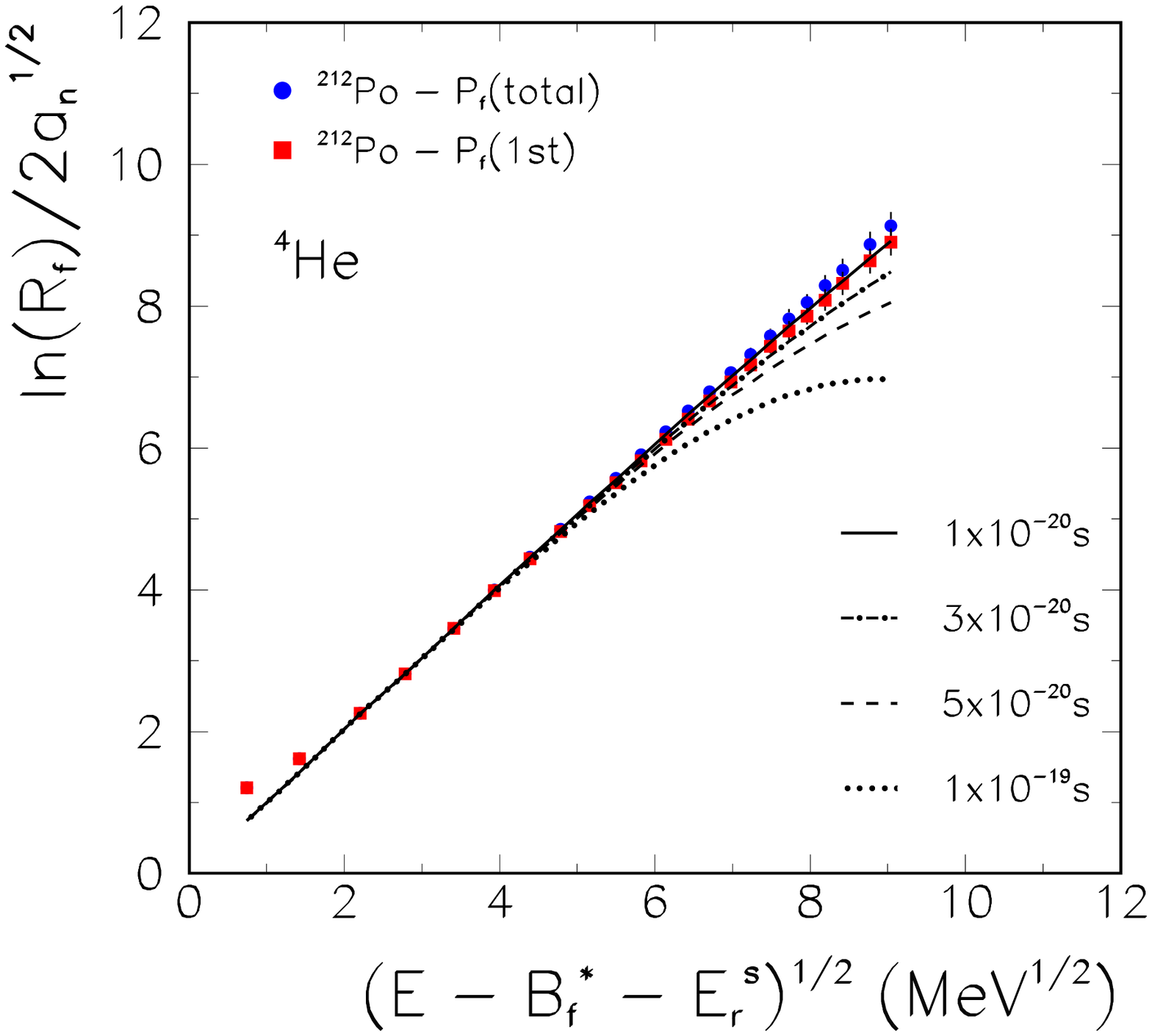,height=7cm}}
 \caption[]{
 {\bf PRELIMINARY.} Same as Fig.~\protect\ref{time} but for first
  chance fission  of $^{212}$Po only.
 }
 \label{rf_first_chance}
\end{figure}

First chance fission at a given excitation energy $E^*$ can be 
determined by subtracting the fission probabilities of two
neighboring isotopes:
\begin{equation}
P_f^{1st} = \frac{\Big(P_f^{tot}(E^*)\Big)_{N,Z} 
            - \Big(P_f^{tot}(E^*-S_n-2T)\Big)_{N-1,Z}}
            {1 - \Big(P_f^{tot}(E^*-S_n-2T)\Big)_{N-1,Z}}.
\label{1stchance_eq}
\end{equation}
Here, $S_n$ represents the separation energy of the last neutron and
$T$ is the temperature of the daughter nucleus given by 
$T=\sqrt{E^*/a_n}$. 
We note that the angular momentum dependence is neglected in this
simple ansatz. The average angular momentum  taken away by one 
neutron can be estimated to be smaller than 0.5$\hbar$.

In Fig.~\ref{1stchance}, we show the preliminary results of this
analysis for $^4$He induced reactions. At excitation energies smaller
than $\sim$45 MeV, 1st chance fission accounts for practically 
all fission events. However, at higher excitation energies,
multi-chance fission sets in and 1st chance fission only accounts 
for $\sim$10\% of the total fission probability at the highest 
excitation energies investigated. It is interesting that 2nd
chance fission becomes even slightly stronger than 1st chance fission 
around 100 MeV.

As pointed out before, the formalism described in Section~\ref{analysis}
has been established for first chance fission only. We  thus 
apply the method to our experimental 1st chance fission results.
In Fig.~\ref{rf_first_chance}, we show the results of this analysis
in comparison with the results using the total fission probability.
Although there is a small difference between the data investigating 
1st chance fission only and those including higher chance fission
at high excitation energies, no significant deviations from the straight
line are visible. Similar results have been obtained for $^3$He 
induced fission. We thus conclude that no deviations from the 
transition state rates have been found and that fission transient 
times must be shorter than 30 zs. It seems likely that that any
excess prescission emission occurs during the descent from 
saddle to scission. If this is the case, then the present fission
results are not in contradiction with recent measurements of prescission
neutron and $\gamma$ rays \cite{Hil92,Pau94,Tho93}.

\section{Summary}
We have measured $^3$He and $^4$He induced fission excitation
functions. The model-independent analysis of these experimental
results allows for the test of the validity of the transition state
rates. It furthermore allows one to determine effective fission barriers
and the ground state shell effects of the compound nuclei investigated.
No deviations from the transition state rate predictions have been
observed in our data. Good agreement between the extracted shell 
corrections and those obtained from the ground state masses have been
found. 
Since the experimental fission rates are well described
by the transition state rates and an upper limit for the fission
transition time of 30 zs could be determined, it seems likely that any 
excess prescission emission occurs during the descent  
from saddle to scission.

\bigskip

{\noindent\bf Acknowledgement}\\
This work was supported by the Director, Office of Energy Research,
Office of High Energy and Nuclear Physics, Nuclear Physics Division
of the US Department of Energy, under contract DE-AC03-76SF00098.

\end{document}